\documentclass[conference]{IEEEtran}
\IEEEoverridecommandlockouts
\usepackage{cite}
\usepackage{amsmath,amssymb,amsfonts}
\usepackage{algorithmic}
\usepackage{graphicx}
\usepackage{textcomp}
\usepackage{xcolor}

\usepackage{booktabs} 
\usepackage{multicol}
\usepackage{multirow}
\usepackage{multirow}
\usepackage{float}  
\usepackage{subfigure}  
\usepackage{adjustbox}

\usepackage{array}
\usepackage{hyperref}

\def\BibTeX{{\rm B\kern-.05em{\sc i\kern-.025em b}\kern-.08em
    T\kern-.1667em\lower.7ex\hbox{E}\kern-.125emX}}
\begin{document}

\title{Multi-Modal Environmental Sensing Based Path Loss Prediction for V2I Communications 

}

\author{
    \IEEEauthorblockN{Kai Wang, Li Yu,  Jianhua Zhang, Yixuan Tian, Eryu Guo, Guangyi Liu*}
    \IEEEauthorblockA{\textit{State Key Laboratory of Networking and Switching Technology, Beijing University of Posts and Telecommunications} \\
    Beijing, China}
    \IEEEauthorblockA{\textit{*China Mobile Research Institution, Beijing, China}}
    \IEEEauthorblockA{Email: \{wangk, li.yu, jhzhang, Tianyx, guoryu\}@bupt.edu.cn}
    \IEEEauthorblockA{*Email: \{Liuguangyi\}@chinamobile.com}
}

\maketitle
 
\begin{abstract}

The stability and reliability of wireless data transmission in vehicular networks face significant challenges due to the high dynamics of path loss caused by the complexity of rapidly changing environments.
This paper proposes a multi-modal environmental sensing-based path loss prediction architecture (MES-PLA) for V2I communications. 
First, we establish a multi-modal environment data and channel joint acquisition platform to generate a spatio-temporally synchronized and aligned dataset of environmental and channel data.
Then we designed a multi-modal feature extraction and fusion network (MFEF-Net) for multi-modal environmental sensing data. MFEF-Net extracts features from RGB images, point cloud data, and GPS information, and integrates them with an attention mechanism to effectively leverage the strengths of each modality.
The simulation results demonstrate that the Root Mean Square Error (RMSE) of MES-PLA is 2.20 dB, indicating a notable improvement in prediction accuracy compared to single-modal sensing data input. Moreover, MES-PLA exhibits enhanced stability under varying illumination conditions compared to single-modal methods.

\end{abstract}


\section{INTRODUCTION}

Advancements in 6G technology will greatly enhance the development of intelligent transportation systems \cite{r8}. In this scenario, vehicle-to-infrastructure (V2I) communication is a crucial technology that can significantly enhance both transportation efficiency and safety. Accurate path loss prediction in V2I communication ensures consistent signal coverage, preventing communication interruptions caused by signal attenuation. This enhances the reliability and security of communication, ultimately improving overall traffic safety. Traditionally, path loss estimation methods can be categorized into statistical and deterministic approaches\cite{3D-MIMO}\cite{r10}\cite{3D-fading-channel-models}. Statistical methods are known for their simplicity and speed; however, they often overlook the environmental factors between transmitters and receivers, resulting in limited reliability, especially in novel or complex environments. Deterministic methods, such as ray tracing, offer high accuracy but come with significant computational overhead, poor real-time performance, and challenges in adapting to dynamically changing environments.

With the rapid advancements in sensing technology and artificial intelligence (AI), path loss prediction based on environmental sensing and AI algorithms offers a promising solution to the challenges mentioned\cite{DTC-6G}\cite{r11}. Sensors enable comprehensive capture of environmental features, while AI excels in processing high-dimensional data and uncovering complex non-linear relationships. By leveraging real-time environmental data, AI models can dynamically adapt to changes, enhancing prediction accuracy and efficiency. This approach demonstrates strong robustness and real-time capabilities, particularly in complex urban communication scenarios.

Some research has explored enhancing path loss prediction by integrating environmental sensing technologies with artificial intelligence algorithms. The authors of \cite{r1} utilized visual data obtained from multi-view sensing cameras to predict path loss. The authors of \cite{r2} utilized side-view images between the transmitter and receiver, along with frequency information, to predict path loss. The work in \cite{r3} employed 2D satellite images of the target area as input to a deep neural network to predict path loss distributions at various UAV altitudes. 
The aforementioned work primarily focused on utilizing images for environmental sensing. Due to the lack of 3D information in images, some research has explored the use of point clouds as a method for environmental sensing. The work in \cite{pc1} extracted features from 3D point clouds and utilized artificial neural networks to improve path loss estimation accuracy. The work in \cite{r5} leveraged Manhattan measurement data at the 28 GHz band, point clouds, and building grid data to predict path loss in urban canyon scenarios.

Environmental sensing has been utilized to some extent in the aforementioned work, but there are still limitations. For example, most approaches tend to rely on a single-modal of sensing, making it exhibit limitations when dealing with dynamic changes in complex scenarios. In some cases, single-modal sensing data is susceptible to interference from external factors such as weather and light conditions, resulting in degraded prediction performance. To address these challenges, we first developed a multi-modal environment data and channel joint acquisition platform for data collection and built a comprehensive environment awareness and channel dataset for urban V2I scenarios. Then we propose MES-PLA, which integrates RGB image, point cloud data, and GPS information, uses a feature extraction module to extract the three modal data respectively, and then uses an attention mechanism to fuse the three features to improve the robustness and accuracy of path loss prediction in complex environments.

\section{DATASET CONSTRUCTION}

\begin{figure}[htbp]
\centerline{\includegraphics[width=0.5\textwidth]{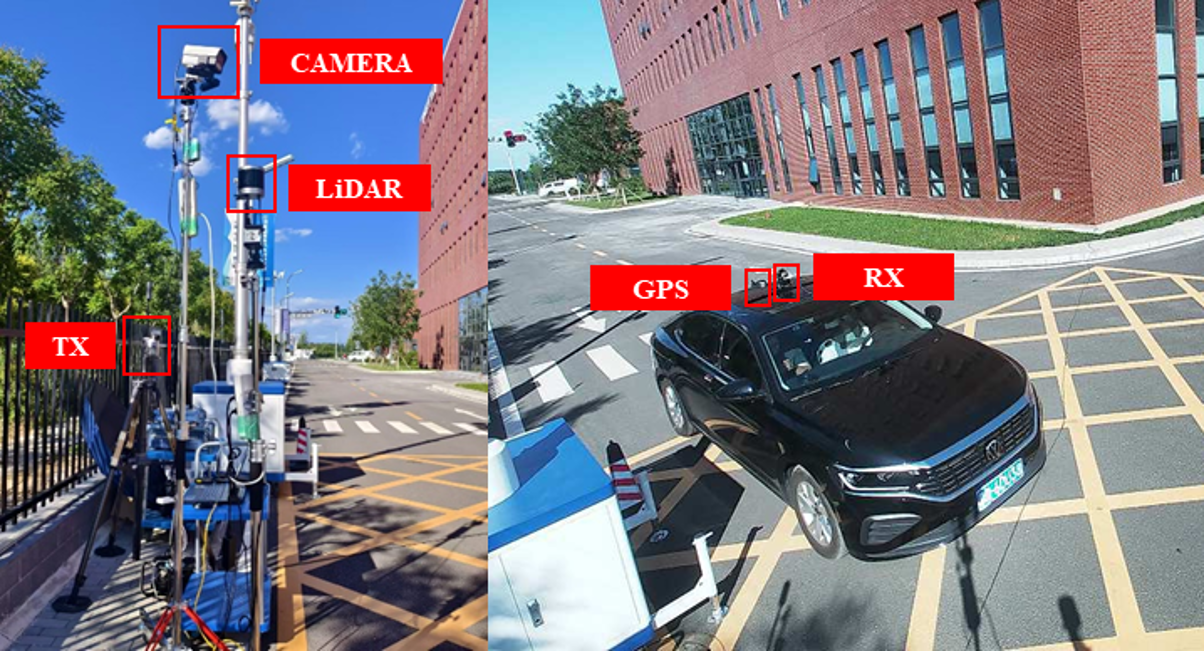}}
\caption{multi-modal environment data and channel joint acquisition platform.}
\label{fig: platform}
\end{figure}

To accurately capture the environmental and channel characteristics in real-world settings, we designed and built a multi-modal environment data and channel joint acquisition platform, as illustrated in Fig. \ref{fig: platform}. This platform is capable of simultaneously collecting environmental data (images, point clouds, GPS) and channel data, enabling the construction of a comprehensive dataset that integrates environmental and channel information. The core components of the platform include LiDAR, two RGB cameras, a GPS receiver,  transmitting and receiving antennas, spectrum analyzers, and computers. Through the coordinated operation of these devices, the platform can simultaneously capture  RGB images, 3D point clouds, positional information, and channel data, enabling precise and synchronized data collection.


The LiDAR (Light Detection and Ranging) captures 3D point cloud data of the surrounding environment. It is mounted at a height of 2.1 meters from the ground to ensure coverage of detailed geometric information from both the street and surrounding buildings. This point cloud data provides spatial layout information of objects in the environment, serving as a key input to the subsequent model. The RGB camera is installed at a height of 3.1 meters and is used to capture 2D RBG images of the scene. These images provide visual features such as texture, color, and edge information. The GPS device is mounted on the roof of the vehicle and provides global positioning data, helping the model account for geographic variations in channel propagation. The transmission antenna is positioned alongside the LiDAR and RGB camera at a height of 1.5 meters, while the receiving antenna is mounted on the vehicle's roof. These antennas collect channel data during signal propagation. The antenna system is synchronized with the environmental data collection devices to ensure precise alignment between the channel and environmental data. To guarantee synchronized data collection, all sensing devices are managed through a unified control platform. During each data collection session, the LiDAR, RGB camera, GPS, and antennas operate simultaneously, ensuring that every frame of data includes complete environmental and channel characteristics.

\begin{figure}[htbp]
\centerline{\includegraphics[width=0.5\textwidth]{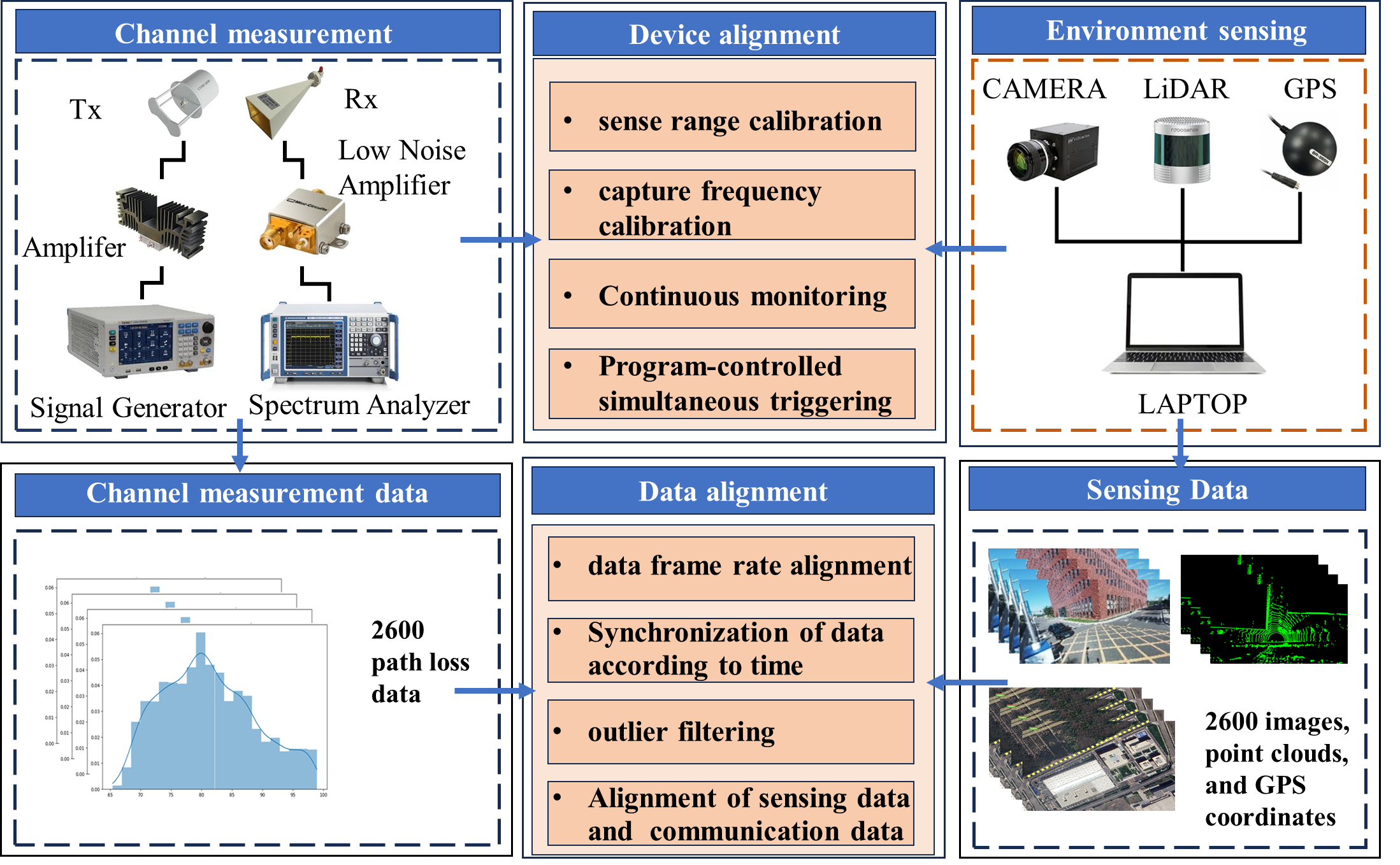}}
\caption{The generation and processing workflow of comprehensive datasets.}
\label{fig:plat_abstract}
\end{figure}

\begin{figure}[htbp]
\centerline{\includegraphics[width=0.5\textwidth]{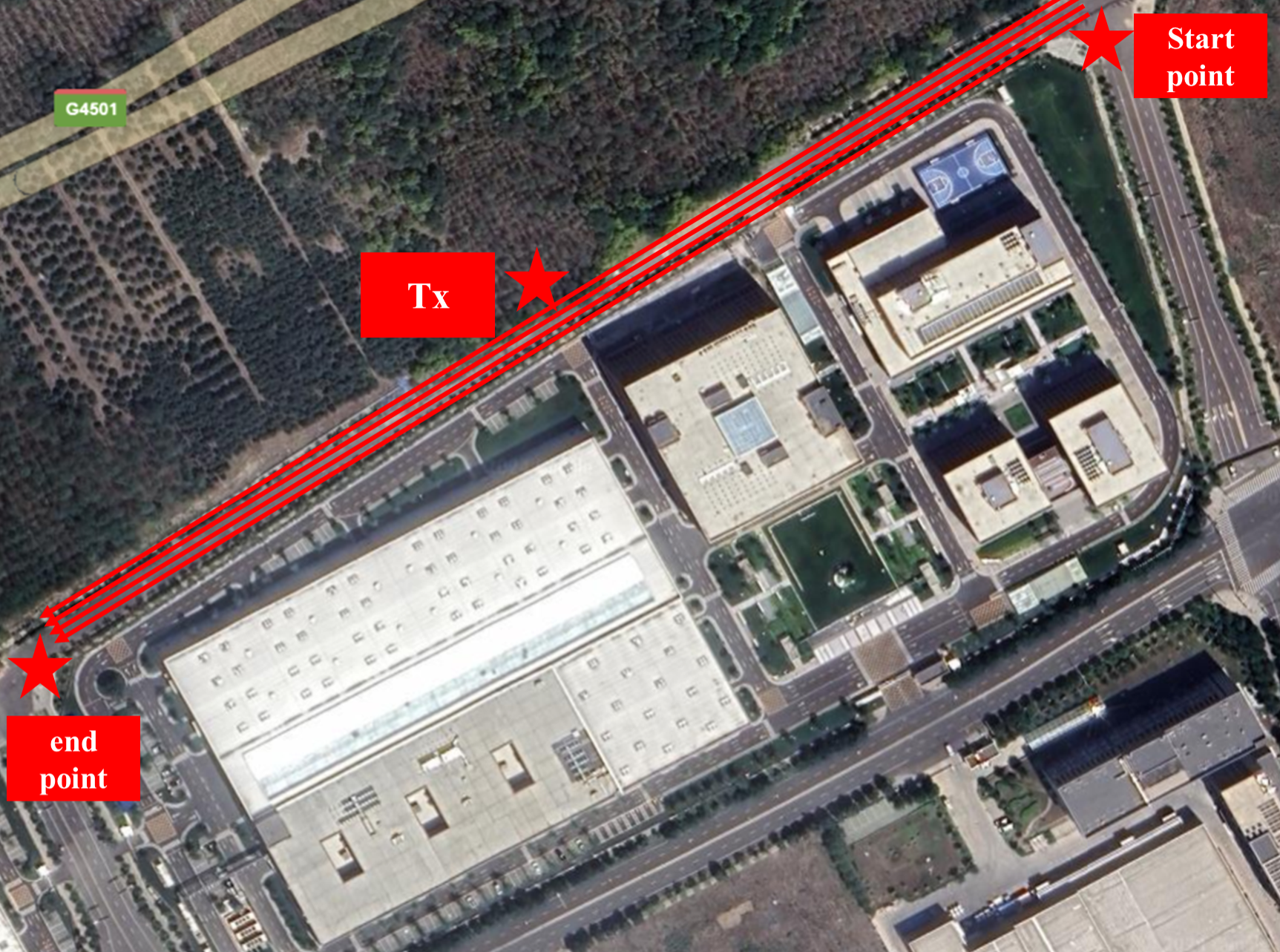}}
\caption{The Rx follows four specific driving routes during the experiment.}
\label{fig: route}
\end{figure}

The selection of the data collection scene is crucial, as it directly affects both the channel propagation characteristics and the environmental complexity. As illustrated in Fig. \ref{fig: route}, we selected a typical urban T-junction street as the data collection site. This street is approximately 14 meters wide and 375 meters long, bordered by five buildings along its length. To ensure the dataset's richness and diversity, we designed four collection routes within this street scene, covering various positions, angles, and signal propagation paths. The driving trajectory of the Rx is represented by the red line. During data collection, the vehicle followed these predefined routes, with each route corresponding to a series of time-synchronized multi-modal data collection points. By conducting multiple trips along these routes, we obtained a sufficient number of multi-modal data samples.

\begin{figure}[t]
\centerline{\includegraphics[width=0.5\textwidth]{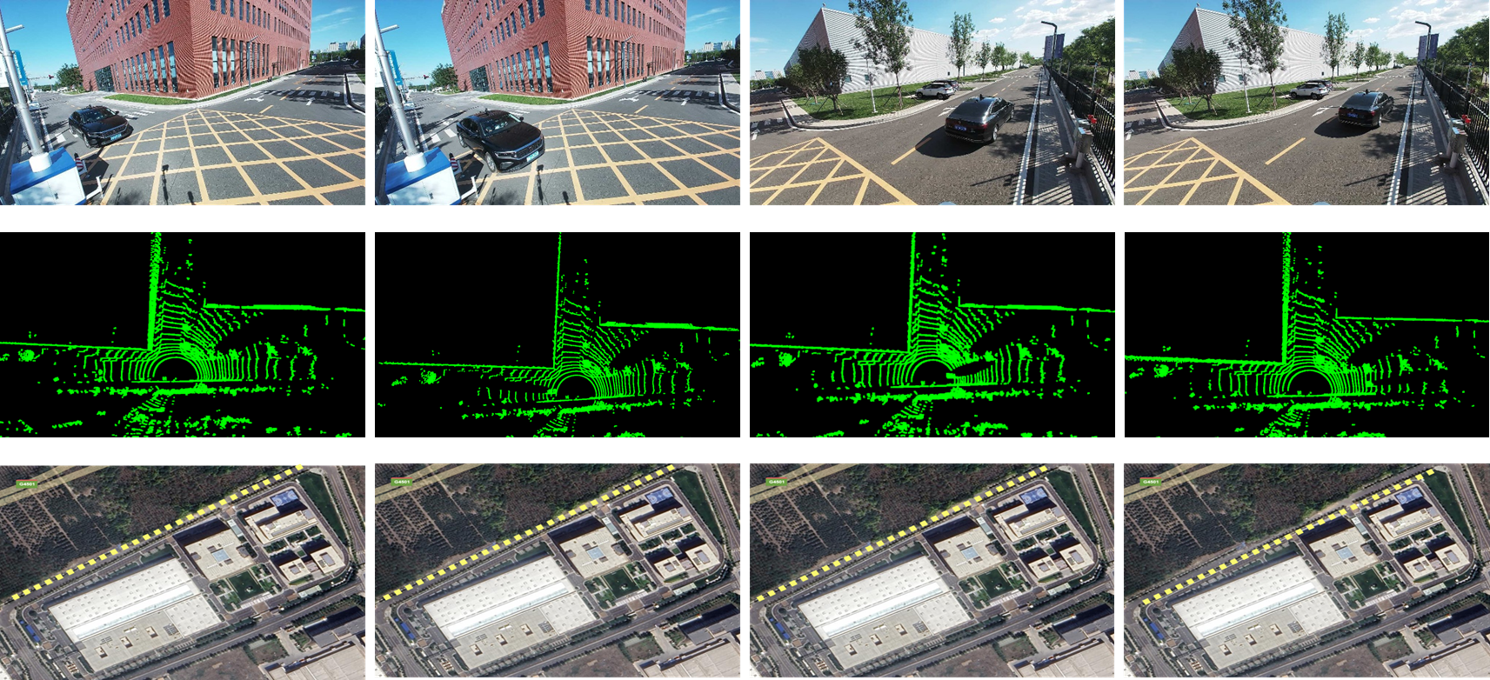}}
\caption{The visualization of the data set, from top to bottom, is RGB image, Lidar point cloud, GPS information, and the yellow dot in the third-row image is the recorded track of GPS.}
\label{fig: data}
\end{figure}

In total, 2,600 data samples were collected from the multi-modal environment data and channel joint acquisition platform in the selected scenario. Each sample consists of one RGB image, one point cloud frame, one GPS coordinate, and one channel path loss measurement. The dataset is presented in Fig. \ref{fig: data}.


\section{METHODOLOGY}
\begin{figure*}[tbp]
\centering
\includegraphics[width=1\textwidth]{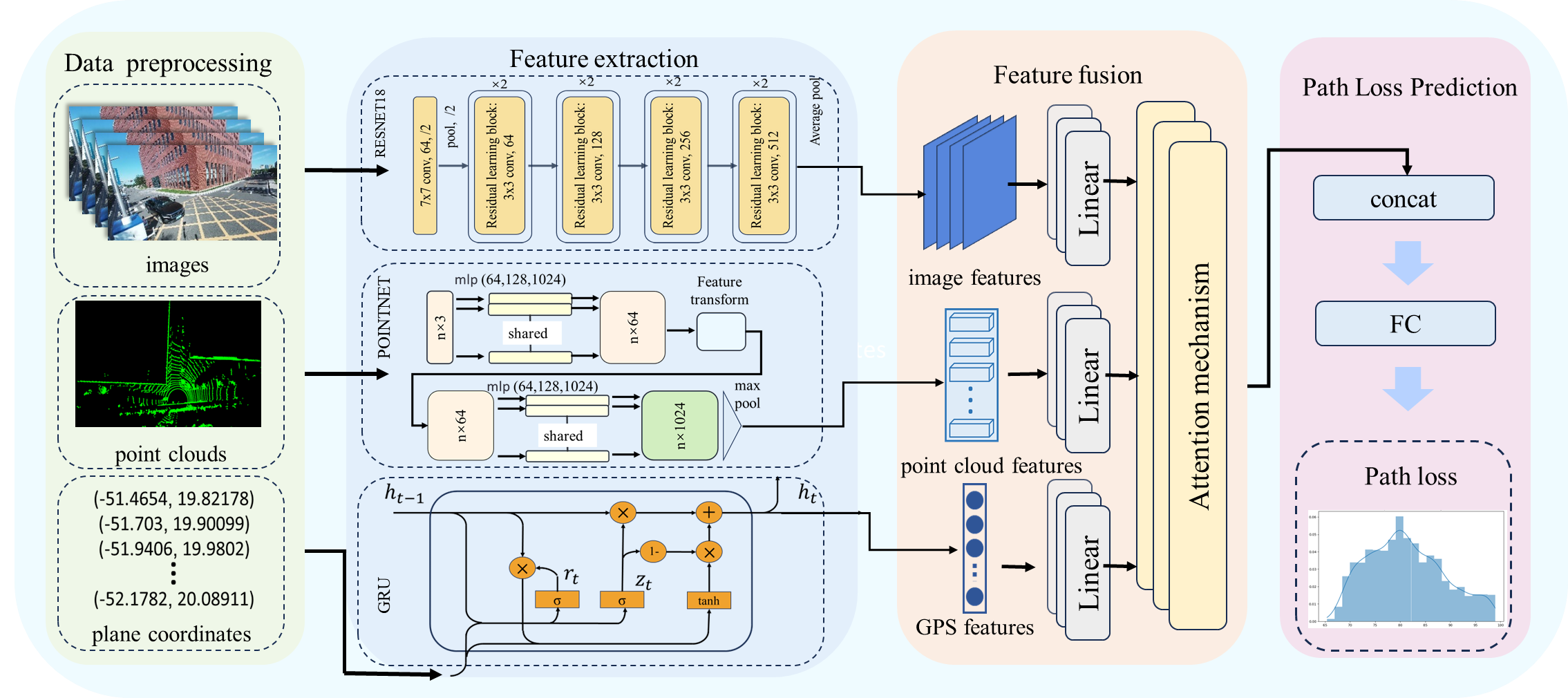}\\
\caption{Architecture forMFEF-Net}
\label{fig: framework}
\end{figure*}

In this section, we will introduce the structure of MFEF-Net, as illustrated in Fig. \ref{fig: framework}. MFEF-Net takes images, point clouds, and GPS information as inputs. Following an initial data preprocessing step, each input is processed by its respective feature extraction network. The extracted features are then integrated using an attention-based fusion mechanism. Finally, the fused features are used to generate the final predictions. This relationship can be represented as
\begin{equation}\label{eq2}
PL \rightarrow f (I_{img}, I_{PC}, I_{GPS})
\end{equation}
where $f$ represents the mapping between the path loss and the input image, the point cloud, and the GPS. $I_{img}, I_{PC}, I_{GPS}$ represent the input image, the point cloud, and the GPS respectively.

\subsection{Data preprocessing}
\subsubsection{RGB image Data Preprocess}




To simulate nighttime conditions, RGB images were transformed into the HSV color space \cite{RGB2HSV}, which separates hue, saturation, and brightness components. This transformation facilitates precise adjustment of brightness levels without affecting other color attributes. The conversion from RGB to HSV is described by the following equations:

\begin{equation}
 V = \max(R, G, B)\label{eq:V} \\
\end{equation}
\begin{equation}
 S = 
\begin{cases} 
0   & \text{, if } V = 0 \\
\dfrac{V - \min(R, G, B)}{V}  & \text{, otherwise}
\end{cases}\label{eq:S} \\
\end{equation}


\begin{align}
H = \begin{cases}
0 & \text{, if } V = \min(R, G, B) \\
60 \times \left( \frac{G - B}{V - \min(R, G, B)} \right) & \text{, if } V = R \\
60 \times \left( 2 + \frac{B - R}{V - \min(R, G, B)} \right) & \text{, if } V = G \\
60 \times \left( 4 + \frac{R - G}{V - \min(R, G, B)} \right) & \text{, if } V = B
\end{cases}
\end{align}

where $ R, G, B \in [0, 1] $ are the normalized red, green, and blue channels. After obtaining the HSV representation, the brightness component $V$ was adjusted by applying a scaling factor $\alpha (0< \alpha<1)$ to reduce the brightness. The modified HSV values were then transformed back into the RGB color space for subsequent analysis, ensuring the simulated images realistically represent nighttime conditions.

\subsubsection{3-D LiDAR Data Preprocess}
LiDAR (Light Detection and Ranging) generates highly accurate 3D point cloud data by emitting laser pulses and capturing their reflections to measure the three-dimensional coordinates of objects. However, environmental conditions often introduce noise points due to insufficient light reflection or surface characteristic variations. To improve reliability and accuracy, it is essential to filter out these outliers from the raw point cloud data. Given the typically large volume of sample points in point cloud data, directly processing them incurs significant computational overhead. To address this problem, a voxel-based downsampling process is employed. This method divides the 3D space into uniformly sized voxel grids, selecting a representative point from each grid to replace all points within it. This approach effectively reduces data volume while preserving the geometric characteristics of the point cloud, thereby lowering computational complexity. Moreover, it retains the spatial structure of the environment, ensuring the accuracy of subsequent path loss prediction tasks. Finally, the processed point cloud data is transformed into a format compatible with the PointNet model for further analysis.

\subsubsection{GPS Data Preprocess}

The raw GPS coordinates are obtained from position sensors, typically represented as latitude and longitude. These coordinates define a specific location on the Earth's surface. However, because the Earth is approximately spherical, directly using latitude and longitude for calculations introduces complexity due to the curvature of the Earth. To simplify computations and analysis, GPS coordinates are converted to a planar coordinate system using the Miller projection method \cite{r14}.
The Miller projection is a map projection technique that transforms spherical latitude and longitude coordinates into planar coordinates. This projection minimizes shape distortion and reduces the exaggeration of areas near the poles. The mathematical transformation is defined as follows:
Let the GPS coordinate of the receiver be denoted as $ G_{Rx}[t]=[lat_{Rx}, lon_{Rx}]$, where $lat_{Rx}$ is the longitude, ranging from $[-180^\circ, 180^\circ]$. $lon_{Rx}$ is the latitude, ranging from $[-90^\circ, 90^\circ]$.
The conversion from Angle to radians is a necessary step because the trigonometric function in the projection formula is measured in radians. The GPS coordinates obtained in radians are expressed as
$ \hat{G_{Rx}[t]}=[\hat{lat_{Rx}}, \hat{lon_{Rx}}]$.
Assuming the central longitude of the projection is $\hat{lat_{Rxo}}$, the projected planar coordinates $P_{Rx}[t]=[X_{Rx}, Y_{Rx}]$ are calculated as:

\begin{equation} \label{eq:MillerX}
X_{Rx} =R_{\mathrm{Earth}} (\hat{lat_{Rx}} - \hat{lat_{Rxo}})
\end{equation}
\begin{equation} \label{eq:MillerY}
 Y_{Rx} = R_{\mathrm{Earth}} \left[ \ln  \tan \left({\pi}/{4} + 0.4  \hat{lon_{Rx}} \right) \right]/0.8
\end{equation}

where $R_{\mathrm{Earth}}$ is the Earth's radius, $\hat{lat_{Rx}}$ and $\hat{lon_{Rx}}$ are the longitude and latitude in radians, respectively. $\hat{lat_{Rxo}}$ is the longitude of the projection’s central meridian, typically set to 0 (the Greenwich prime meridian). The scaling factor for latitude in the Miller projection is 0.8 to mitigate the excessive stretching of areas at high latitudes. Using the above formulas, latitude and longitude are systematically transformed into planar coordinates $P_{Rx}[t]=[X_{Rx}, Y_{Rx}]$.

\subsection{Feature extraction and fusion}
We construct a feature fusion network based on multi-modal environmental sensing data input, which aims to predict the path loss of the wireless channel. The input of the network included three modalities: image, point cloud, and location information (GPS), and the feature extraction network was designed to extract the deep features of each modality. Finally, the attention mechanism was used to fuse the features of the three modalities to predict the path loss.

We adopt the classical ResNet-18 \cite{r15} architecture in the image feature extraction stage as the feature extractor. ResNet-18 consists of four stages, each of which contains two basic blocks, each of which consists of two 3×3 convolutional layers and a shortcut connection. Each stage progressively reduces the spatial resolution and increases the channel dimensions through convolution and maximum pooling operations, with the dimensions being 64, 128, 256, and 512, respectively. We adapt ResNet-18 to retain its backbone and remove the last fully connected part, and output to a 512-dimensional feature feature vector to form high-level features of image modalities.

The point cloud feature extraction module, based on the PointNet architecture \cite{r16}, integrates the Spatial Transformation Network (STN), point-wise convolution, and global pooling operations to efficiently extract global features. Initially, the input point cloud data is filtered and downsampled. The STN module then aligns the 3D coordinates by learning a rotation and translation transformation matrix, mitigating the effects of rotation, scale, and deformation on feature extraction and enhancing model robustness. Next, the point cloud passes through three layers of one-dimensional convolution with output channels of 64, 128, and 1024, progressively capturing features from low to high levels. Each convolution layer is followed by Batch Normalization and the ReLU activation function, which improve training stability and enhance the network's non-linear representational capabilities. Maximum pooling extracts the most significant features on the point dimension, and obtains a global feature vector of 1024 dimensions, representing the features of the entire point cloud. This architecture effectively addresses the irregularity and sparsity of point clouds. Employing lightweight point-wise convolution significantly reduces computational complexity while providing robust feature representation for efficient point cloud processing and subsequent tasks.


Location information is used for feature extraction using a Gated Recurrent Unit (GRU) \cite{r17}. GRU is an efficient recurrent neural network suitable for modeling the temporal dependence of sequential data. In this module, we represent the input position information as a 3D vector $(x, y, z)$ and input it into the single-layer GRU for processing. The dimension of the hidden layer of GRU is set to 128, and the output is the hidden state of the last time step, which is used to characterize the global features of position information. 

We employ an attention mechanism \cite{r12} to fuse important features from image, point cloud, and GPS features in multi-modal data. Features from these three modalities are concatenated and the attention mechanism is applied to weigh them accordingly. In this case, the attention mechanism calculates the importance of each fused feature, which can be seen as generating a weighted feature representation. The attention weights are learned dynamically, thus enabling the model to focus on the most relevant parts of the input data. Most of the attention mechanism in this model generates attention weights by applying a linear layer to the connected feature vectors. These weights are then normalized using a softmax function to ensure that they sum to 1, thus providing a relative importance for each feature. The weighted sum of the features is then used as input to the final prediction layer. The features are then weighted and combined through a full connection layer to predict channel path losses.

\section{SIMULATION RESULTS AND ANALYSIS}

\subsection{Metrics}
This paper selects root mean square error (RMSE) as the primary evaluation metric. This metric effectively reflects the differences between the model's predicted results and the actual values, thereby assisting in the analysis of the model's accuracy and reliability.
The formula of RMSE is defined as 
\begin{equation}\label{eq1}
\mathit{RMSE = \sqrt{\frac{1}{N_{test}} \sum_{i=1}^{N_{test}} (PL_{pred}^{(i)} - \hat{PL}_{true}^{(i)})^2}}
\end{equation}


$N_{test}$ is the total number of dataset; $PL_{true}$ is the measured true value of channel path loss.
$PL_{pred}$ is the path loss value predicted by the model.

\begin{table}[ht]
    \centering
    \renewcommand{\arraystretch}{1.2} 
    \caption{PERFORMANCE OF DIFFERENT MODAL INPUTS AND DIFFERENT LIGHT CONDITIONS}
    \label{tab:example}
    \begin{tabular}{|>{\centering\arraybackslash}m{4.5cm}|>{\centering\arraybackslash}m{1.5cm}|>{\centering\arraybackslash}m{1.5cm}|}
        \hline
        \multirow{2}{*}{\textbf{Input Modal}} & \multicolumn{2}{c|}{\textbf{RMSE(dB) of predication}} \\
        \cline{2-3}
        & \textbf{daytime} & \textbf{nighttime} \\
        \hline
        RGB Images & 7.05 & 9.05 \\
        Point Clouds & 8.59 & $\backslash$ \\
        RGB Images \& GPS & 2.63 & 6.25 \\
        RGB Images \& Point Clouds & 2.57 & 3.04 \\
        Point Clouds \& GPS & 3.19 & $\backslash$ \\
        RGB Images \& Point Clouds \& GPS & 2.20 & 2.44 \\
        \hline
    \end{tabular}
\end{table}

\begin{figure}[htbp]
\centerline{\includegraphics[width=0.5\textwidth]{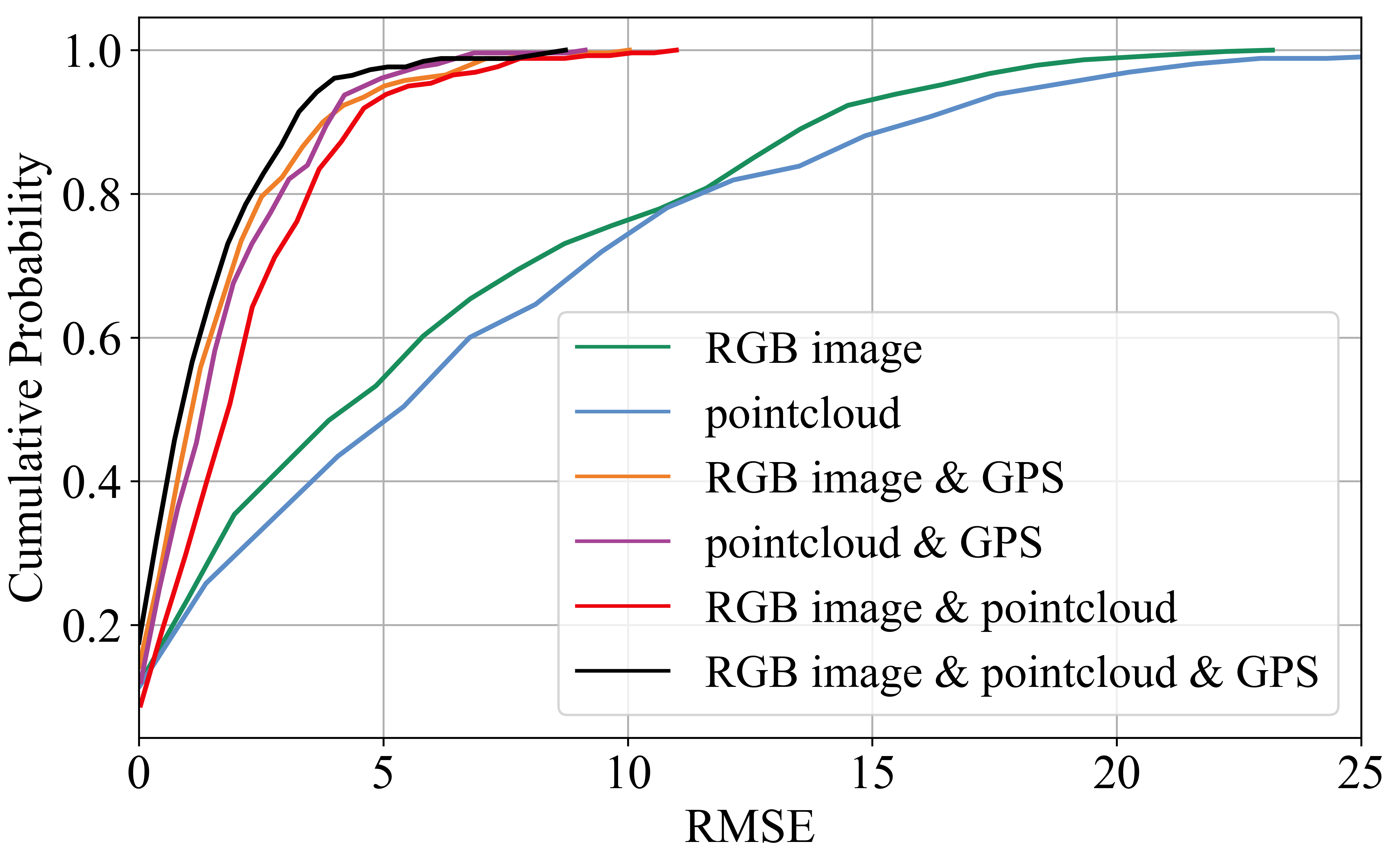}}
\caption{CDF of different modal combinations.}
\label{fig: CDF}
\end{figure}

Fig. \ref{fig: CDF} illustrates the cumulative distribution function (CDF) comparing the performance of different modalities. The X-axis represents the RMSE, which quantifies the average magnitude of the prediction error. When the RMSE threshold is low, the multi-modal curves rise more steeply and reach a higher CDF value, indicating superior prediction performance compared to the single-modal method. Among all multi-modal approaches, MES-PLA achieves the best performance.
MES-PLA can process data from different modes to better cope with complex and dynamic environmental conditions. Specifically, image-based modalities are highly sensitive to fine textures and variations in scene lighting. However, when lighting conditions are inadequate or rapidly changing, image-based perception can be significantly impaired, potentially reducing predictive performance. In contrast, point cloud data generated by LiDAR is unaffected by lighting conditions, providing stable 3D structural information even in low-light environments. Thus, when environmental conditions exceed the adaptability of a single modality, MES-PLA allows the system to leverage the strengths of each modality, compensating for the limitations of others and enhancing the robustness and generalization of the model.

\begin{figure}[htbp]
\centerline{\includegraphics[width=0.5\textwidth]{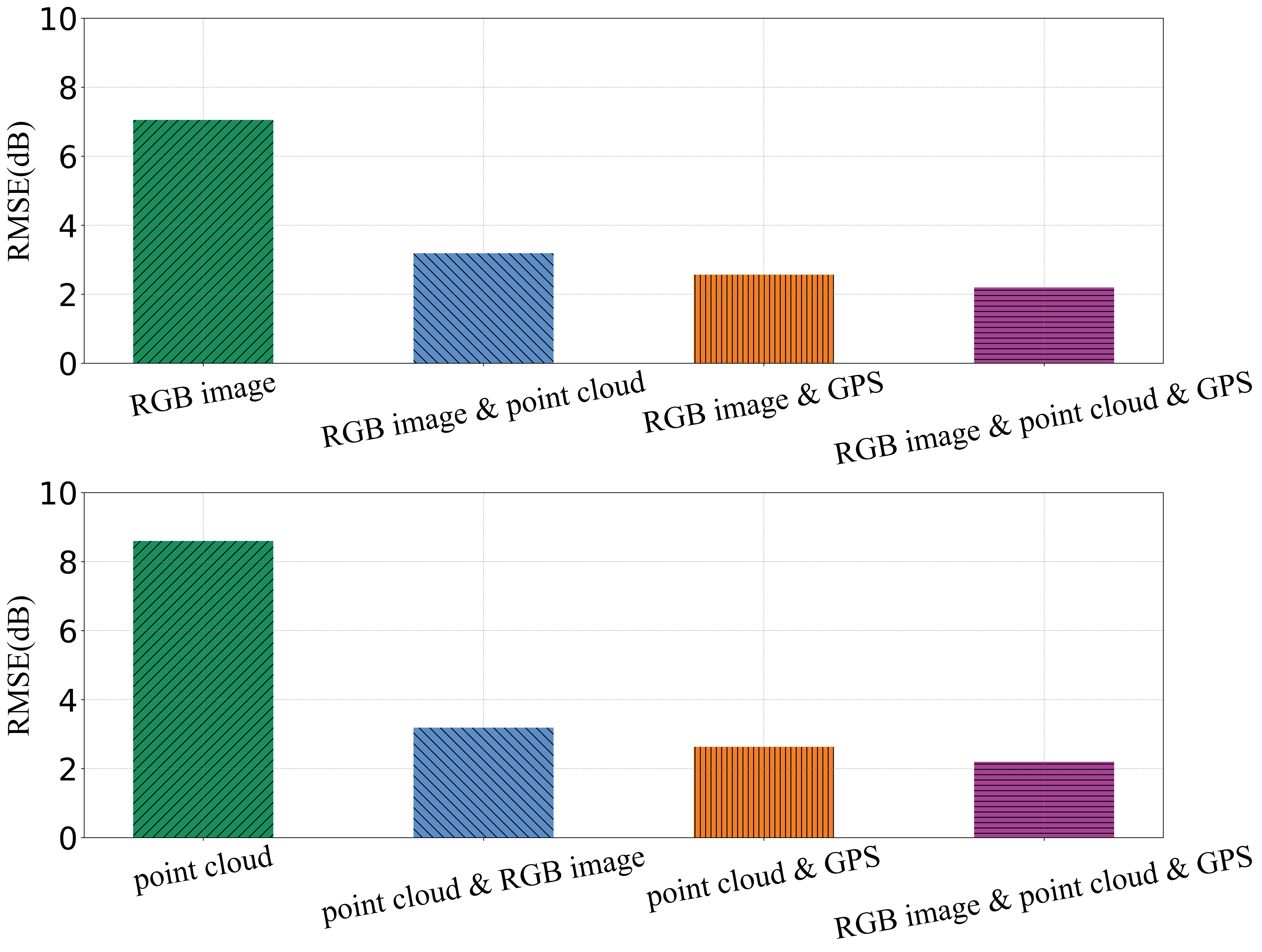}}
\caption{RMSE comparison between single-modal and multi-modal methods.}
\label{fig: RMSE}
\end{figure}

Fig. \ref{fig: RMSE} shows the RMSE when using the point cloud and RGB image as the base input, respectively, and adding other modal data. For example, when using pure image input, RMSE was reduced by 54.75\% when adding point cloud data and 63.59\% when adding GPS data. When all three modal data are used as inputs, RMSE was 68.75\% lower than single-modal image input. For point cloud input only, the three combinations decreased by 62.80\%, 70.55\%, and 74.42\%, respectively.

\begin{figure}[htbp]
\centering
\subfigure[]{
\includegraphics[width=0.48\textwidth]{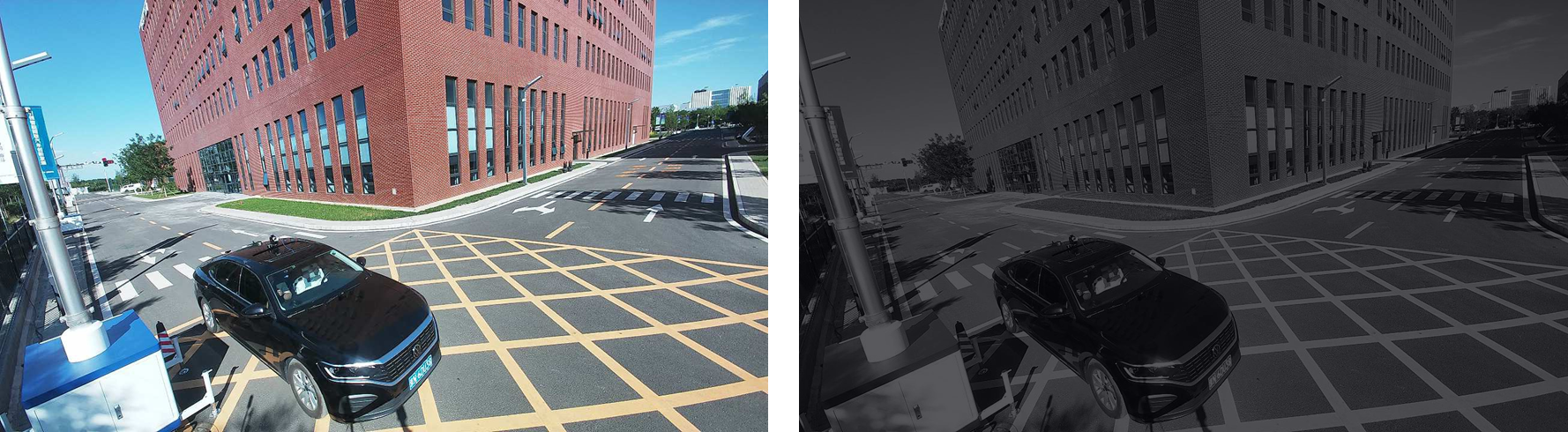}
}
\quad
\subfigure[]{
\includegraphics[width=0.5\textwidth]{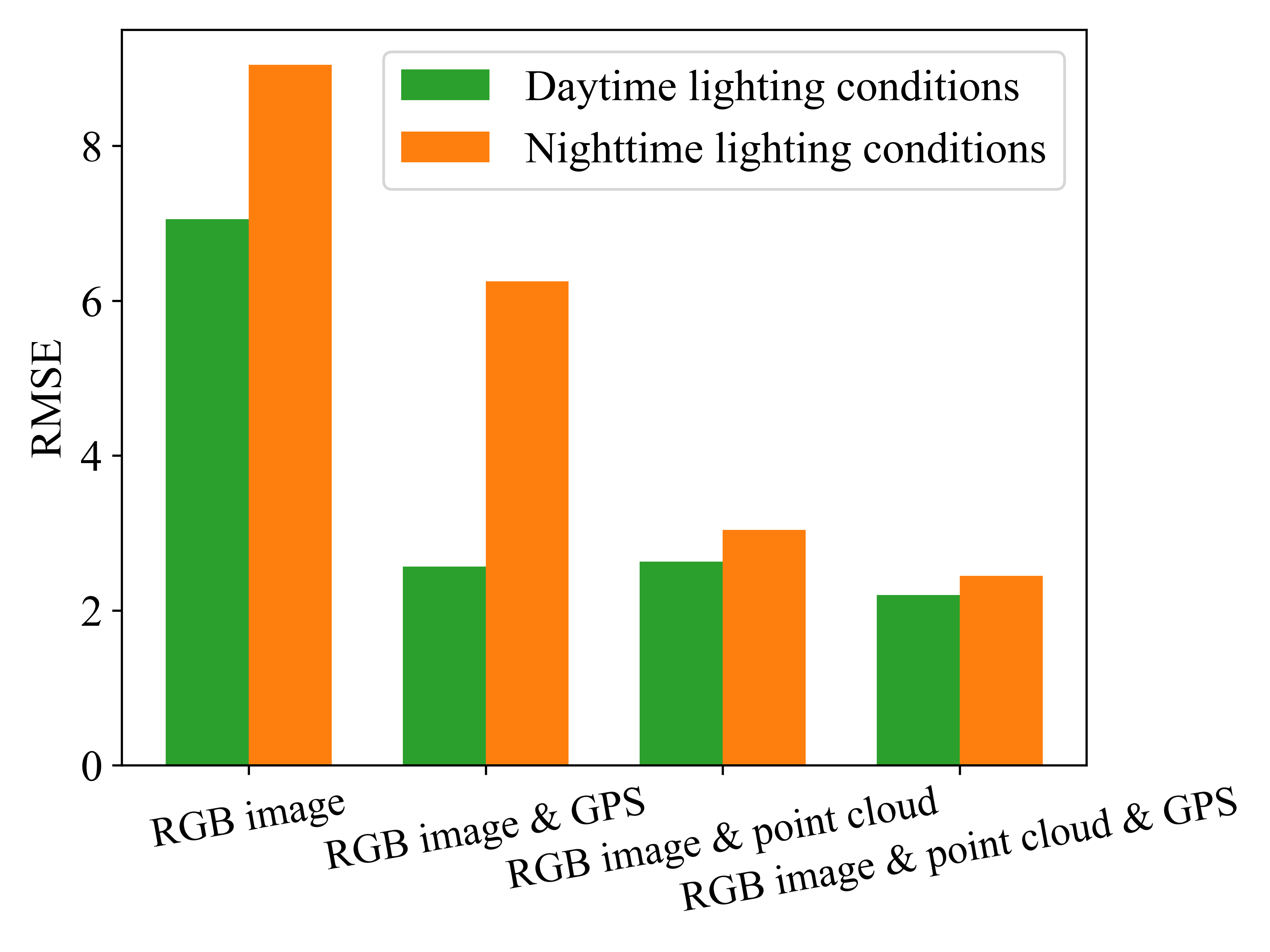}
}

\caption{Simulation results of the model under varying lighting conditions. (a)Image data under different lighting conditions. (b) RMSE comparison between normal light and night simulation.}

\label{fig: light}
\end{figure}

The influence of varying lighting conditions (day vs. night) on the performance of the path loss prediction model is explored. To simulate typical day and night lighting environments, the brightness of RGB images is adjusted accordingly. Since LiDAR is unaffected by lighting conditions, the point cloud data remains unchanged. In the daytime simulation, the average brightness of the image was 135.24 candela per square meter (cd/m²), consistent with the brightness level typically observed in clear daylight or bright light conditions. In the nighttime simulation, the average brightness was reduced to 18.95 cd/m², reflecting the typical lighting conditions of an urban street at night, particularly in areas with inadequate artificial lighting. The simulation results, shown in Fig. \ref{fig: light}, indicate that the RMSE increased by 28.3\% when using single-modal image input for prediction, and by 143.4\% when combining image and GPS data. The other two combinations, including point cloud inputs, showed RMSE increases of 15.4\% and 11.2\%, respectively. These results demonstrate that the performance of MES-PLA is more stable than that of the single-modal method under low-light conditions, with a smaller impact on prediction accuracy. GPS data provides crucial spatial information, such as the relative distance between the transmitter and receiver. This distance is closely linked to path loss, making it essential for accurate channel modeling and path loss prediction.

\section{CONCLUSION}
In this paper, we propose MES-PLA, a path loss prediction method based on multi-modal environmental sensing that uses RGB images, point cloud data, and GPS data as inputs, with path loss as the output. MES-PLA leverages the strengths of different sensors, enabling the method to effectively handle complex and dynamic channel propagation environments. Simulation results demonstrate that MES-PLA achieves higher accuracy than single-modal methods and maintains stable performance under varying lighting conditions. GPS data plays a crucial role in improving prediction accuracy. Future work will focus on optimizing the model's efficiency and incorporating additional environmental factors to support path loss predictions in more complex urban scenarios.

\section*{Acknowledgment}

This work is supported by the National Key R\&D Program of China (Grant No. 2023YFB2904805), the National Natural Science Foundation of China (No. 62401084), and BUPT-CMCC Joint Innovation Center.




\bibliographystyle{IEEEtran}
\bibliography{reference}

\end{document}